\documentclass[aps,twocolumn,showpacs]{revtex4}

\usepackage{hyperref,graphicx}

\begin{document}

\title{\bf Efficient sampling in complex materials at finite temperature:
the thermodynamically-weighted activation-relaxation technique}

\author{Normand Mousseau}
\email{normand.mousseau@umontreal.ca}
\affiliation{
        D\'epartement de Physique,
	Universit\'e de Montr\'eal,
        C.P. 6128, Succursale Centre-ville Montr\'eal,
        Qu\'ebec, Canada H3C 3J7.}

\author{G.T. Barkema}
\email{barkema@phys.uu.nl}
\affiliation{
	Institute for Theoretical Physics, 
        Utrecht University, Leuvenlaan 4, 3584 CE Utrecht, 
        the Netherlands}

\date{\today}

\begin{abstract}
We present an accelerated algorithm that samples correctly the
thermodynamic ensemble in complex systems where the dynamics is controlled
by activation barriers. The efficiency of the thermodynamically-weighted
activation-relaxation technique (THWART) is many orders of magnitude
greater than standard molecular dynamics, even at room temperature and
above, in systems as complex as proteins and amorphous silicon.

\end{abstract}


\maketitle

Throughout physics, chemistry and biology, a large proportion
of atomistic processes take place on time scales many orders of
magnitude longer than the typical phonon period. These processes
are out of reach of single time-scale algorithms, such as molecular
dynamics (MD), which can reach simulation times equivalent to the
microsecond at best. In view of this limitation, considerable effort
has been devoted in the last few years to develop algorithms that
allow spanning multiple time scales. These algorithms include the
activation-relaxation technique (ART)~\cite{barkema96,malek00} and similar
techniques~\cite{doye97,henkelman99} which sample the configurational
landscape by identifying transition paths from minimum to minimum,
as well as accelerated schemes based on molecular dynamics such as
hyper-MD~\cite{voter97}, temperature-assisted dynamics~\cite{voter97b}
and others~\cite{laio03}.

ART and similar methods have been applied with success to study
the topology of the energy landscape and activated mechanisms
in a wide range of materials including amorphous and crystalline
semiconductors~\cite{barkema98,elmellouhi03,middleton01}, glassy
materials~\cite{mousseau00a}, atomic clusters~\cite{doye99a,malek00}
and proteins~\cite{malek01,mortenson01,wei02}.  ART defines events in
the energy landscape as a two-step process: (1) the system is first
activated from a local energy minimum to a nearby saddle-point and (2)
then relaxed to a new minimum. Since the landscape constructed by ART
consists only of local minima connected via saddle points, the entropic
contributions to sampling are neglected so it is not possible to guarantee
a proper statistical sampling, especially at elevated temperatures where
the harmonic approximation breaks down.

Accelerated MD schemes, on the other hand, rely either on deforming
the energy landscape~\cite{voter97,laio03} or on some projection from
high-temperature simulations~\cite{voter97b}. In the first case, the
energy basin surrounding a local minimum is filled according to various
rules. In the second case, multiple simulations are performed at high
temperature and as soon as an event occurs, the time is rescaled with an
appropriate factor. These methods work only for simple systems, however,
where it is possible to have a detailed {\it a priori} knowledge of the
parameters defining the landscape. As soon as a wide range of barriers
come into play, such as in proteins or disordered materials, it becomes
very difficult to apply the appropriate transformations in order to
sample correctly and efficiently the phase space.

In this Letter, we present an algorithm that provides a proper statistical
sampling of the energy landscape at a wide range of temperatures
irrespective of the complexity of the landscape.  Combining molecular
dynamics with ART, the thermodynamically-weighted activation-relaxation
technique (THWART) samples the thermodynamically relevant parts of
the phase space, hopping over barriers that can be many times higher
than $k_BT$.

\begin{figure}
\includegraphics[width=6cm]{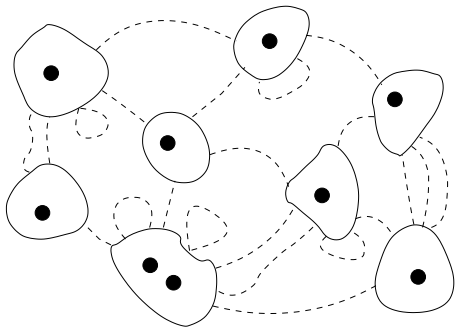}
\caption{Sketch of a two-dimensional energy landscape.  The black dots
  denote the locations of local-energy minima. These minima are part of
  basins, bounded by a line of constant lowest-curvature (solid line);
  the percolating region surrounding the basins is called the saddle
  region.  Basin-to-basin trajectories as generated by THWART are
  indicated by dashed lines.  Constrained to ensure detailed balance,
  the trajectories come back to were they started if they fail to
  find a boundary. 
\label{fig:sketch}}
\end{figure}

THWART generates motion in the energy landscape, a 3N-dimensional
hypersurface, on which the height is given by the potential energy of a
configuration as a function of its $3N$ atomic coordinates, where $N$ is
the number of atoms in the system.  At low temperature, a configuration
typically spends most of its time oscillating thermally around a local
minimum, hopping over an energy barrier only when a thermal fluctuation
transfers large amounts of energy onto a single mode.  In this regime,
it is possible to separate the energy landscape into two types of
regions: basins and saddle regions. The {\it basins} are regions around
local-energy minima where all Hessian eigenvalues, corresponding to the
curvature of the landscape, are above a threshold value $\lambda_0$.
The {\it saddle regions} have at least one direction with a curvature
(eigenvalue of the Hessian matrix) below this threshold value. These
regions surround metastable points such a first or higher-order saddle
points.  A two-dimensional sketch of an energy landscape, paved according
to these criteria, is shown in Fig.~\ref{fig:sketch}.

At low temperature, the equilibrium properties of a system are
determined by the basins, where the Boltzmann weight $\exp(-E/k_BT)$
is large and most of the sampling takes place.  The partition function,
$Z=\int d\vec{X} \exp(-E(\vec{X})/k_BT)$, is thus well approximated by
integrating only over the basins. The basins, however, are disconnected
regions in phase space and a proper sampling requires integrating over
many of these basins.  To ensure efficient sampling, it is therefore
essential to accelerate the rate at which basins are visited.

THWART achieves this acceleration in two steps. In the basin regions,
configurations are evolved using standard MD. We select the NVT ensemble
and use the velocity Verlet algorithm for the integration. During the
MD simulation,  we monitor the lowest eigenvalue of the Hessian which
defines the position of the boundary separating the basin from the saddle
region. As soon as this lowest eigenvalue reaches a given threshold
$\lambda_0$, the MD is suspended and the hopping phase starts. The
atomic positions at the basin boundary is identified by $\vec{x}_0$
and the velocities by $\vec{v}_0$.

From $\vec{x}_0$, on the basin boundary, we construct a fully reversible
path that leads to a new basin, crossing an energy barrier. For this,
we follow the eigendirection corresponding to the lowest eigenvalue away
from the basin until the eigenvalue crosses the threshold $\lambda_0$
from below. In order to ensure detailed balance, the trajectory into
the saddle region is further constrained to move on a hyperplane with
constant potential energy. Specifically, the activated trajectory is
generated by iterating the following equation:
\begin{equation}\label{eq:act}
  \vec{x}_{i+1} = \vec{x}_i + \frac{\Delta t}{2}
  \left(\vec{h}_i+\vec{h}_{i+1}\right) + c
  \left(\vec{F}_{\perp;i}+\vec{F}_{\perp;i+1}\right).
\label{eq:follow}
\end{equation}
Here, $\vec{h}_i$ is the normalized eigenvector at $\vec{x}_i$
corresponding to the lowest (most negative) Hessian eigenvalue,
$\vec{F}_{\perp;i}$ is the component of the force at $\vec{x}_i$
perpendicular to $\vec{h}_i$, $\Delta t$ is a constant factor that
determines the size of the increment, and $c$ is a multiplicative
constant, chosen to project the trajectory onto the hyperplane of constant
potential energy.  The orientation of $\vec{h}_0$ is chosen initially so
that it points towards the direction of more negative curvature, i.e.,
away from the initial basin; it is updated at each step by requiring
that the inner product of the local eigenvector $\vec{h}_i$ with that
at the previous step, $\vec{h}_{i-1}$ be always positive.

The move along the eigendirection corresponding to the lowest eigenvalue
changes the total configurational energy.  The last term on the right-hand
side of Eq.~(\ref{eq:follow}) corrects for this change and constrains the
trajectory to the hyperplane with constant energy.  Because the initial
configuration is thermalized, its configurational energy is well above
that of the local minimum, with roughly $k_BT/2$ of additional potential
energy per degree of freedom. By transferring energy to a specific degree
of freedom from the heat bath formed by all the other $3N-1$ degrees of
freedom, the algorithm reproduces therefore the statistical mechanism
responsible for crossing barriers.  Simply moving along the force,
to enforce this constraint tends to bring the configuration back to its
original position. Instead, we correct the energy by moving in the reduced
space generated by the hyperplane perpendicular to this eigendirection,
projecting the force onto this hyperplane, $\vec{F}_{\perp;i}$.

Eq.~(\ref{eq:follow}) is iterated until the lowest eigenvalue passes
the threshold (from below, this time) and the configuration reaches
the boundary of a basin with position $\vec{x}_p$. At this point, the
activation phase is stopped and the MD is resumed with the new positions
and velocities $\vec{v}_p=\vec{v}_0$.

The path generated from $\vec{x}_0$ to $\vec{x}_p$ is fully reversible:
a configuration in basin $p$ reaching $\vec{x}_p$ would trigger the
activation, bringing it to the other end of this path, in $\vec{x}_0$.
Reversibility is ensured by the symmetric criterion for entering
and leaving the saddle region as well as by keeping the path on a
hyperplane of constant energy. Reversibility is not sufficient for
detailed balance, however.  In continuous space, it is also necessary
to verify that an infinitesimal volume surrounding $\vec{x}_0$ remains
constant as the configuration moves along the path leading to $\vec{x}_p$.
This transformation, given by the Jacobian, determines the change in phase
space, or entropy, between these two points. We have verified numerically
that the Jacobian of transformation is unity along the whole path from
one basin boundary to the other. In combination with reversibility, the
preservation of phase space ensures that detailed balance is respected.

We also note that the path does not always lead to a new basin.  In our
simulations, it is not rare to see the trajectory form a circular path,
coming back exactly at the initial point, $\vec{x}_0$, on the boundary
after a long excursion in the saddle region. If the path does not close
on itself, it generally connects to a different basin.


Having established the validity of the algorithm, we apply THWART to two
non-trivial systems: amorphous silicon and a small peptide.  Simulations
on amorphous silicon are performed on models with two different sizes
and initial configurations. The 1000-atom model was generated with ART
nouveau~\cite{malek00} using a modified Stillinger-Weber (mSW) potential
fitted to the amorphous phase~\cite{vink01a}. This configuration is
well-relaxed and is described in Ref.~\cite{valiquette03}. The 500-atom
model was produced with a bond-switching algorithm~\cite{highq,vink01b}
and simply relaxed with the mSW potential, showing higher strain than
the 1000-atom model.

\begin{figure}
\includegraphics[width=8cm]{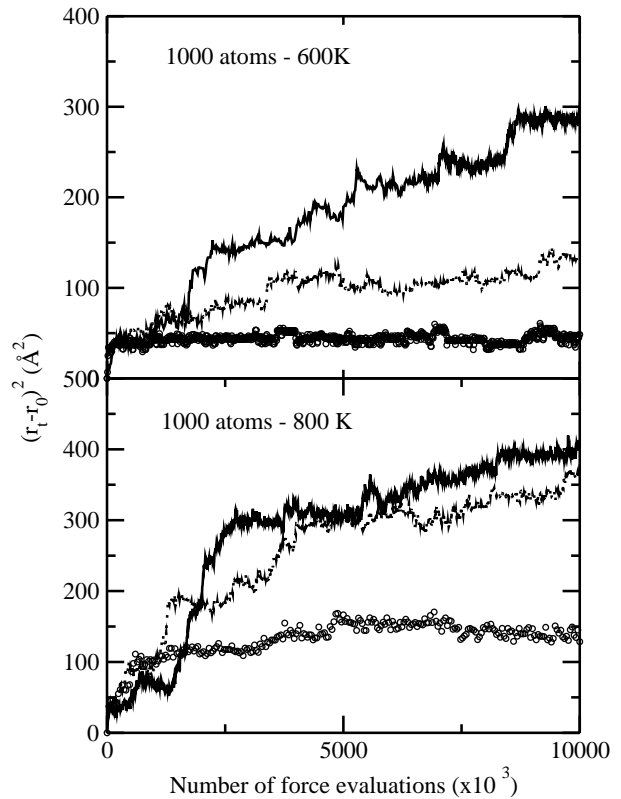}
\caption{Total squared displacement as a function of the number of force
  evaluations, at temperatures 600 K (upper box) and 800 K (lower box),
  obtained with MD (circles) and THWART simulations (lines).
  For the THWART simulations at 600 K, we use $\lambda_0
  = -5$\AA$^2$ (solid line) and -15 \AA$^2$ (dotted line). At 800 K, we
  use $\lambda_0 = -5$\AA$^2$ (solid line) and -10 \AA$^2$ (dotted line).
\label{fig:asi1000}}
\end{figure}

\begin{figure}
\includegraphics[width=8cm]{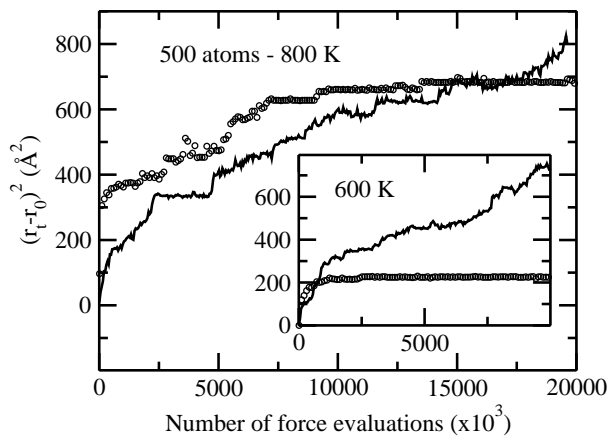}
\caption{Squared displacement as a function of the number of force
  evaluations, at a temperature of 800 K, obtained with MD (circles)
  and THWART simulations (lines) with $\lambda_0 = -20$\AA$^2$. The 
  inset shows the same at 600 K, and with $\lambda_0 = -5$\AA$^2$.
\label{fig:asi500}}
\end{figure}

Both models were then evolved with MD in the standard NVT ensemble
at a temperature of 600 K and 800 K, i.e., well below the melting
transition temperature of $\sim$2000 K for this potential, but near the
crystallization temperature from the amorphous phase. The MD simulations
are integrated with time steps of 1 fs and run over a simulated time
of 10 ns, except for the 500-atom configuration at 800 K, which is
run for 20 ns.  Figs.~\ref{fig:asi1000} and \ref{fig:asi500} show the
total squared displacement, $\langle r^2 \rangle$, between the initial
configuration (which is energy-minimized) and the quenched configuration
at time $t$, measured in the number of force evaluations (with 1 force
evaluation per time step), for these four distinct runs.

For the well-relaxed 1000-atom model at 600 K, the figure shows that after
a few initial local rearrangements, the MD runs remain trapped around
a few nearby minima for more than 9 ns. We find a similar situation for
the 800 K simulation, with the configuration diffusing slightly further
but still failing to explore more than a few nearby basins during the
10 ns simulation.

The situation is quite different with THWART: In the basin, we follow the
same MD procedure as above, while computing the lowest eigenvalue every 50
steps using a 20-level Lanczos scheme. The MD run is continued until this
lowest eigenvalue falls below the threshold value $\lambda_0$.  To avoid
moving back and forth along the same path, the MD procedure is required to
take at least 400 steps.  From this point, we apply Eq.~(\ref{eq:follow})
until the lowest eigenvalue increases above the same threshold.
The total squared displacement measured as a function of the number of
force evaluations for THWART is also plotted in Figs.~\ref{fig:asi1000}
and ~\ref{fig:asi500}; 10 million force evaluations correspond roughly to
5000 events.  As can be seen, THWART explores the energy landscape many
orders of magnitude faster than MD at the same temperature.  THWART is
slightly slower at short times, because it performs one event at a
time, while MD can activate multiples events in parallel across the
model. However, it rapidly surpasses MD. In particular, THWART does not
seem to become trapped either at 600 K or 800 K. It is to be expected,
however, that at high temperatures, when the number of events occurring
in parallel starts proliferating, MD will become more efficient than
THWART. From our simulations, this should happen close to the melting
point.

The efficiency of THWART depends on the value of the threshold, the
sole parameter in THWART.  Its value impacts the ratio of basin to
saddle regions on the energy landscape; a very low value of $\lambda_0$
reduces THWART to MD, while a high value makes THWART more similar to ART.
The impact of the threshold is shown in Fig.~\ref{fig:asi1000}, which
plots the total squared displacement as a function of force evaluation
for two values of $\lambda_0$ at 600 K and 800 K.  At both temperatures,
THWART diffuses faster with a higher threshold, which reduces the basin
size but also delivers a higher fraction of open paths.

Even though the 500-atom model is more strained than the 1000-atom one,
the configuration gets trapped very rapidly at 600 K with MD, within
less than 1 ns. At 800 K, however, the configuration is able to overcome
many barriers and diffuse significantly, becoming trapped only after
about 10 ns of simulation (see Fig.~\ref{fig:asi500}).  In spite of
this considerable collective relaxation, THWART overcomes MD at 800 K,
demonstrating the general efficiency of the method.


Results on a small 10-residue peptide show that THWART is also
more efficient than MD for this system. We use an artificial
peptide of sequence AAAAAGAAAA with interactions described by
CHARMM19~\cite{charmm19} and an ASP solvation term~\cite{asp} as
implemented by Ponder in his program Tinker~\cite{tinker}.  The peptide
is first relaxed near its energy minimum using THWART. This sequence
possesses a large number of metastable minima surrounding this
lowest-energy state.

\begin{figure}
\includegraphics[width=4cm]{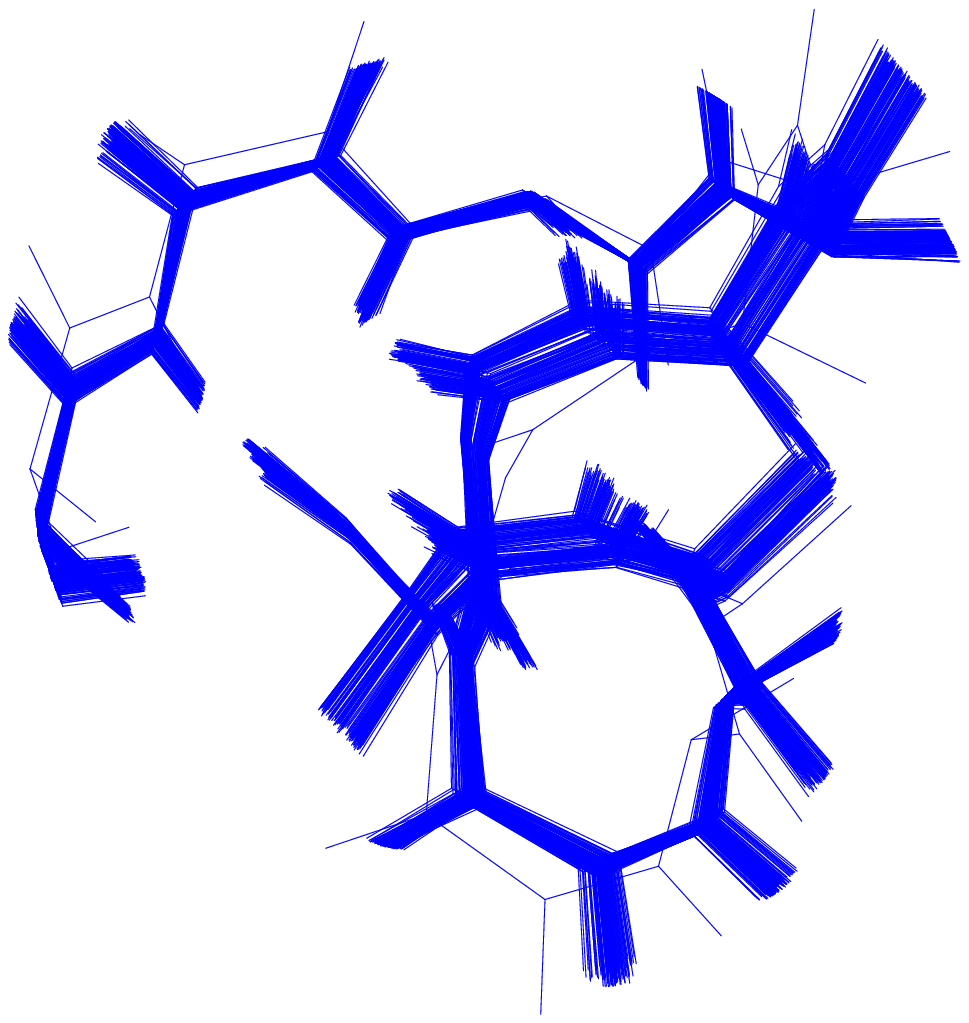}
\includegraphics[width=4cm]{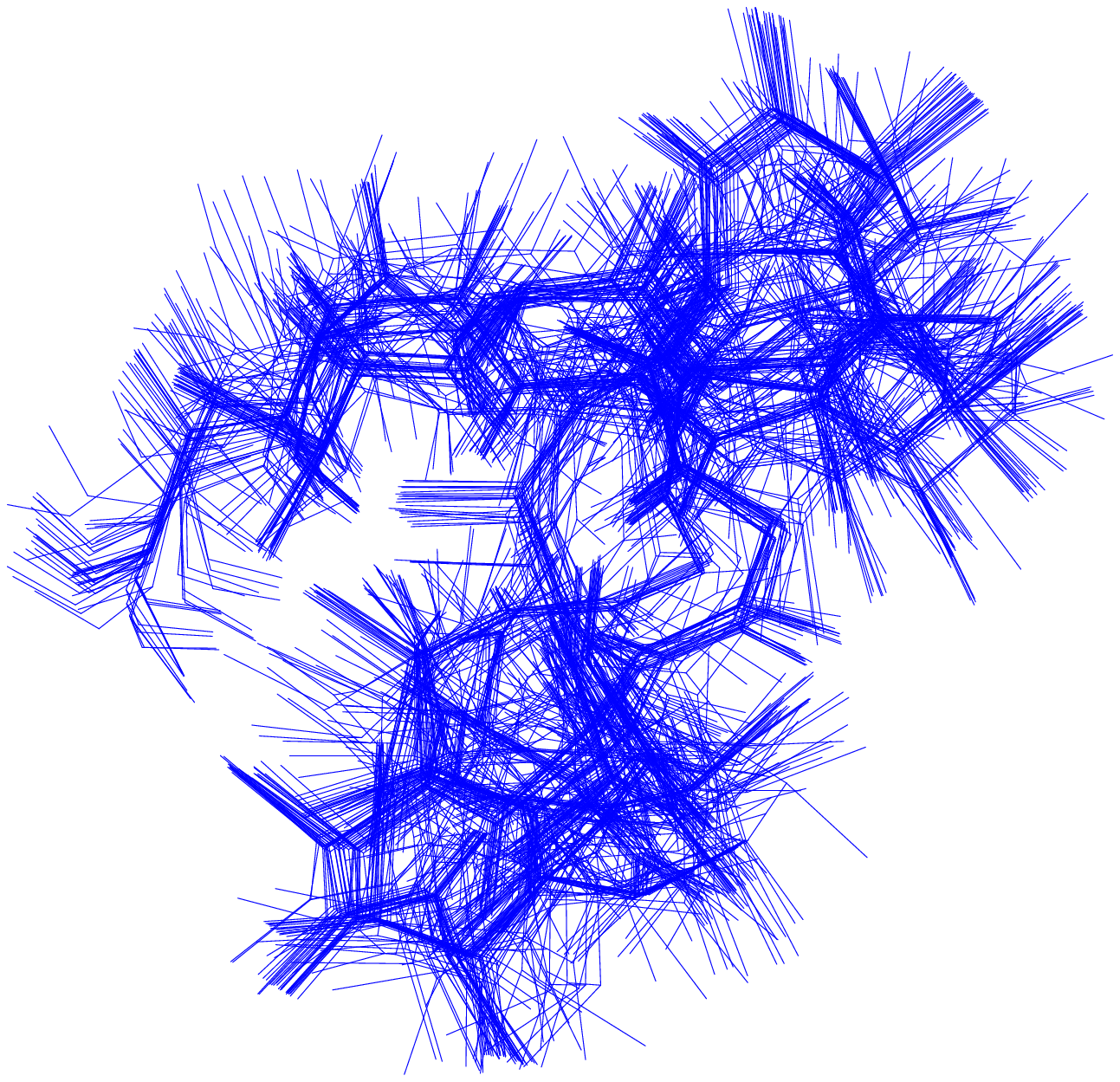}
\caption{Superimposed energy-minimized backbone configurations of the
  peptide at different times, in MD (left) and THWART (right)
  simulations at a temperature of 600K. About 100 configurations are
  superimposed for each method. 
\label{fig:peptide}}
\end{figure}

We first perform a 300-ps constant-temperature MD simulation using
Tinker. To characterize the efficiency of sampling, the energy-minimized
conformations after every 2 ps are graphically superimposed in the left
panel of Fig.~\ref{fig:peptide}. Clearly, the conformation is trapped
in a few nearby local minima. Next, a THWART simulation is performed of
the same system over 300 000 force evaluations, or about 150 events. The
energy-minimized conformations are superimposed in the right panel of
Fig.~\ref{fig:peptide}.  THWART clearly samples the phase space around
the initial state much more efficiently than MD.

In summary, the thermodynamically-weighted activation-relaxation technique
is an accelerated algorithm for sampling efficiently the relevant parts of
the phase space in systems where the dynamics is controlled by activation
barriers.  THWART samples correctly the thermodynamic ensemble, with
an efficiency many orders of magnitude greater than standard molecular
dynamics, even at room temperature and above. We have shown that THWART
is efficient in various complex systems such as proteins and amorphous
silicon. THWART should be particularly useful to study materials such
as glasses and proteins in water, where other accelerated techniques fail.

{\it Acknowledgements.} NM acknowledges support from the Natural Science
and Engineering Council of Canada as well as NATEQ. NM is a Cottrell
Scholar of the Research Corporation.


\end{document}